\documentclass[prl,twocolumn,superscriptaddress,tightenlines]{revtex4-1}
\usepackage{bibunits}
\usepackage{amsmath,amssymb,bm}
\usepackage{graphicx}
\usepackage{epstopdf}
\usepackage{latexsym}
\usepackage{subfigure}
\usepackage{color}
\usepackage{natbib}
\usepackage{hyperref}
\usepackage{braket}
\hypersetup{
  colorlinks,
  citecolor=magenta,
  linkcolor=blue,
  urlcolor=blue}

\def\beq{\begin{equation}}
\def\eeq{\end{equation}}
\def\beqr{\begin{eqnarray}}
\def\eeqr{\end{eqnarray}}
\def\bK{{\mathbf{K}}}
\def\bk{{\mathbf{k}}}
\def\bq{{\mathbf{q}}}
\def\bp{{\mathbf{p}}}
\def\cG{{\cal G}}
\newcommand{\E}{\ensuremath{\mathrm{e}}}
\newcommand{\I}{\ensuremath{\mathrm{i}}}

\begin{document}

\title{Interaction induced Dirac fermions from quadratic band touching
  in  bilayer graphene }
\author{Sumiran Pujari}
\address{Department of Physics \& Astronomy, University of
  Kentucky, Lexington, KY-40506-0055}
\author{Thomas C. Lang}
\address{Institute for Theoretical Physics, University of Innsbruck, 6020 Innsbruck, Austria}
\author{Ganpathy Murthy}
\address{Department of Physics \& Astronomy, University of Kentucky, Lexington, KY-40506-0055}
\author{Ribhu K. Kaul}
\address{Department of Physics \& Astronomy, University of Kentucky, Lexington, KY-40506-0055}
\begin{abstract}
We revisit the effect of local interactions on the 
quadratic band touching (QBT) of the Bernal
honeycomb bilayer model using renormalization group (RG)
arguments and quantum Monte Carlo (QMC) simulations. We
present an RG argument which predicts, contrary to previous studies, that
weak interactions do not flow to strong coupling even if the free
dispersion has a QBT. Instead they
generate a linear term in the dispersion, which causes the
interactions to flow back to weak coupling.  
Consistent with this RG scenario, in unbiased QMC
simulations of the Hubbard model
we find compelling evidence that antiferromagnetism turns on at a finite
$U/t$, despite the $U=0$ hopping problem having a QBT. The onset of antiferromagnetism takes place at a continuous
transition which is consistent with
2+1 d Gross-Neveu criticality.  We conclude that generically in models of bilayer
graphene, even if the free dispersion has a QBT, small local interactions generate a Dirac
phase with no symmetry breaking and that there is
a finite-coupling transition out of this phase to
a symmetry-broken state.
\end{abstract}
\maketitle

\begin{bibunit}[apsrev]

The interplay of band topology and interactions
has taken center stage in condensed matter physics. The interest in
band topology was rekindled by the discovery of graphene, where two bands
touch at linearly dispersing Dirac points~\cite{CastroNeto09}. Short range interactions are known
to be irrelevant at the Dirac fixed point, making the linear band
touching a stable many-body phenomenon for weak interactions. Strong interactions trigger an
instability to a broken symmetry state~\cite{herbut2006:graphene,Meng10:hch,Sorella12:hch}.  

Subsequently it was pointed
out that bilayer graphene harbors a quadratic band
touching (QBT) at half-filling in the nearest neighbor hopping
model~\cite{mccann2006:bil}. Renormalization group studies have shown that interactions
are marginally relevant at a QBT~\cite{sun2009:qbt}. This led to
predictions of the stabilization of a
symmetry broken state even for arbitrarily weak short-range
interactions in bilayer graphene~\cite{vafek2010:bil1,vafek2010:bil2} in contrast to
single layer graphene. The
nature of symmetry broken state depends on the form of the
interactions and a plethora of phases have been proposed (see 
e.g.~\cite{vafek2010:bil1,vafek2010:bil2,zhang2010:bil,nandkishore2010:bil,lemonik2010:bil}). 
The Bernal honeycomb bilayer (BHB) lattice with Hubbard interactions
was studied as an example of this kind of ordering~\cite{lang2012:bil}. Numerical studies
using quantum Monte Carlo concluded that this model was N\'eel ordered for all values of the
coupling $U/t$ consistent with the predictions of weak-coupling~\cite{vafek2010:bil2} and
functional renormalization group (RG) \cite{lang2012:bil}.

\begin{figure}[!t]
\centerline{\includegraphics[angle=0,width=\columnwidth]{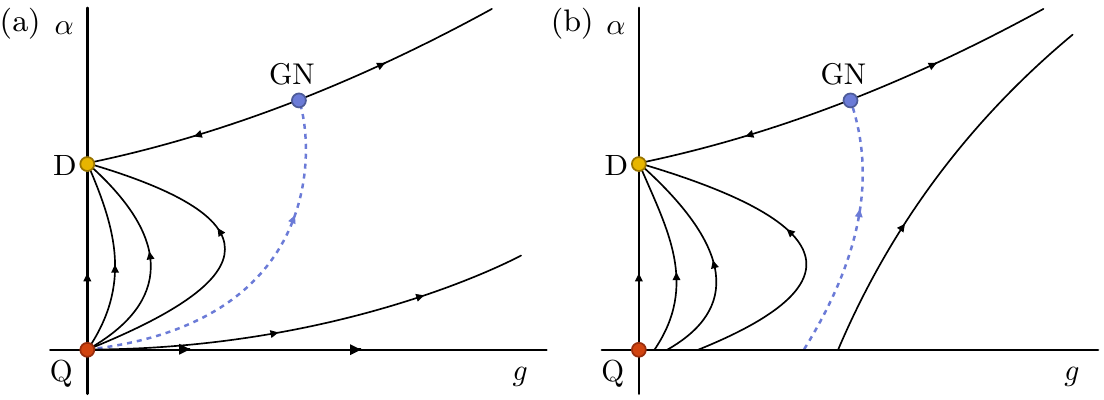}}
\caption{Two schematics for the renormalization group flows for the Bernal stacked bilayer in
  the space of $\alpha$ (linear term in the
  dispersion) and $g$ (generic quartic interactions). [c.f. Eq.~(\ref{eq:lw})
  for definitions]. The fixed points are
  Q: quadratic band touching. D: Dirac linear dispersion. GN:
  Gross-Neveu. (a) The RG flow implied by previous studies. In this
  scenario, weak interactions always flow to strong
  coupling when one begins with a QBT, i.e. along $\alpha=0$. 
(b) RG flow advocated for in this work, when $\alpha$ arises in an RG
flow due to interactions. Note that crucially in (b), $\alpha$ is generated even
  when one begins on the $\alpha=0$ line. The point marked as
  D indicates one of a set of fixed points with $\alpha\neq 0$ which all have a
  linear Dirac dispersion at low energy, but different details of the
  electronic structure. 
}
\label{fig:quad_rg}
\end{figure}

On inclusion of certain symmetry preserving hoppings beyond the
shortest range on the BHB (called trigonal warping), a linear term in the
dispersion is
present at low energies resulting in Dirac nodes~\cite{mccann2006:bil}.  It has been generally assumed that if the
trigonal warping terms
are very weak microscopically, for practical purposes they can
be neglected -- they only cause the symmetry breaking instability to appear
at a small interaction strength controlled by their size
\cite{Cvetkovic2012:trigwarping}, leading to the RG flow in
Fig.~\ref{fig:quad_rg}(a). Here we
present arguments and evidence that neglecting linear terms in the
dispersion is not justified in determining the fate of short range
interactions at the quadratic band touching in bilayer graphene. Even
in a model BHB system where the trigonal
warping terms are microscopically zero and there is a true QBT in the
free problem, interactions generate the linear
term in an RG flow. At weak coupling, we thus conclude
that interactions cause the emergence of a
Dirac phase from a QBT in bilayer graphene. The
instability to a symmetry broken ordered state then takes place at a finite-strength of electron-electron interactions. [c.f. Fig.~\ref{fig:quad_rg}(b)]. Field theoretic
arguments suggest that this transition, if continuous, would be of the Gross-Neveu (GN)
type~\cite{herbut2006:graphene}. As a concrete test of our RG scenario we present QMC simulations
of the Hubbard interaction on the bilayer graphene lattice. We show
that despite the model having a QBT at $U=0$, interactions cause
N\'eel ordering only at a finite value of $U_c\approx 2.6$. The phase transition between
the non-N\'eel and N\'eel phase is fully consistent with the GN
universality class, consistent with Fig.~\ref{fig:quad_rg}(b).

{\em RG argument:} 
In the absence of spin-orbit coupling, symmetries guarantee that two spin
degenerate low energy bands touch at the ${\bf K}$ and ${\bf K^\prime}$ points. Expanding around
the ${\bf K}$ point, a symmetry
based study \cite{mccann2006:bil,vafek2010:bil2} of allowed dispersions for the part quadratic in the fermionic operators $\Psi$ of the
Hamiltonian density has the following form up to second order in ${\bf k}$,
\begin{eqnarray}
\label{eq:lw}
h^{(2)}({\bf K}+{\bf k}) &=& \Psi^\dagger_{{\bf K} +{\bf k}} \left ( {\rm Re}[\phi({\bf k})]\sigma^x+ {\rm Im}[\phi({\bf k})]\sigma^y
            \right ) \Psi _{{\bf K}+{\bf k}},\nonumber \\
 \phi({\bf k}) &\equiv&\alpha (k_x + \I k_y) +\beta (k_x - \I k_y)^2
                   +{\cal O} (k^3).
\end{eqnarray}
The crucial point for our study is that generically a linear term $\alpha$ is
allowed by symmetries. At quartic order, $H^{(4)}$ has nine independent
symmetry allowed terms of the form $\Psi^\dagger \Psi^\dagger \Psi \Psi$ with coupling constants $g_i$ ($i \in [1, 9]$)~\cite{vafek2010:bil2}. 

\begin{figure}[!b]
\centerline{\includegraphics[angle=0,width=\columnwidth]{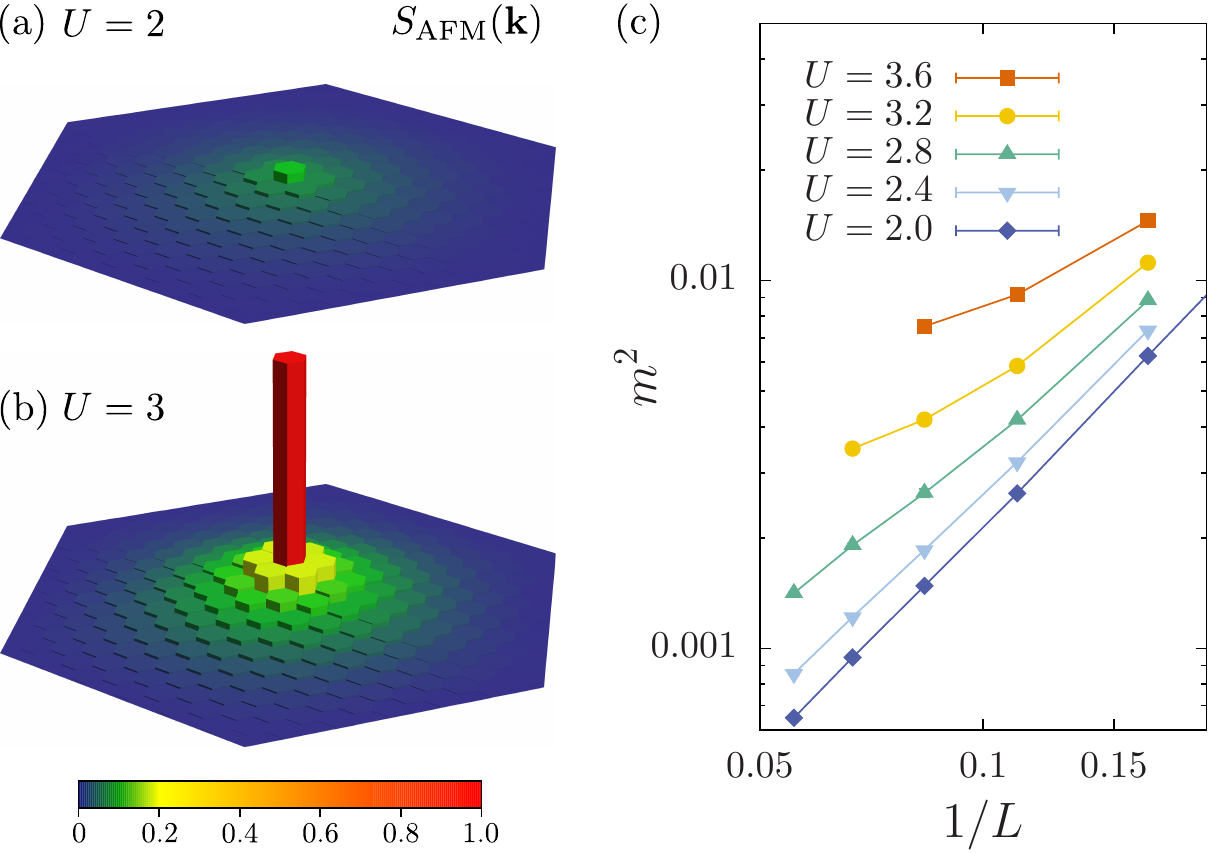}}
\caption{Structure factor $S_{\rm AFM}({\bf k})$ and order parameter $m^2$
  data for $H_{\rm BSB}$ with $t=t_\perp=1$ (fixed throughout the
  paper). In (a) and (b) we show the structure factor in ${\bf
    k}$-space for an ${L=18}$ lattice and its peak at ${\bf \Gamma}$. Note the dramatic
  appearance of a Bragg peak structure as one changes from $U=2$ to $U=3$,
  indicative of a phase transition. (c) Finite size scaling of the
  order parameter $m^2\equiv S_{\rm AFM}({\bf \Gamma})/L^2$, showing a
  finite $m^2$ for $U=3.6$ and 3.2, and then a dramatic suppression for smaller $U$. }
\label{fig:msq}
\end{figure}

From the RG point of view the quadratic band touching 
is a fixed point, called Q ($\alpha, g_i=0$), which has $z=2$. Power counting shows
$\alpha$ is relevant at Q and the
$g_i$ are marginal -- all other couplings are irrelevant~\cite{bernal2016ph:fn}.
A one-loop calculation finds that $g_i$ are
generically marginally relevant~\cite{vafek2010:bil1,vafek2010:bil2}; Q
is hence a multicritical point in the RG sense. We now ask how Q is
affected when we add 
only $g_i$ but no $\alpha$ (this corresponds to adding interactions to
a free dispersion with a true QBT). Wilsonian RG tells us
that all terms that are allowed by symmetries will appear
as we integrate high energy degrees of freedom. Since no symmetry
forces $\alpha=0$ in the dispersion for the Bernal stacked
bilayer, it must be generated. Since at Q the linear term is relevant and the interactions are only
marginal, at weak coupling the linear term will always grow faster and hence the RG
flows will take us to a fixed point with Dirac fermions (see SM for
more details). At this fixed
point called D, interactions are 
irrelevant and hence it is a stable phase
of matter (just like in single layer graphene). Once the $g_i$ cross a finite order-one strength they
will eventually win over the generated $\alpha$ and a flow to strong
coupling will ensue. The separatrix between the flow to weak and
strong coupling can flow to an intermediate coupling fixed point, the
GN universality class --
Fig.~\ref{fig:quad_rg}(b). We note that in models with higher
symmetry than the BHB, the QBT can be symmetry protected~\cite{sun2009:qbt,murray2014:qshqbt}, then weak
interactions will flow to strong coupling --
Fig.~\ref{fig:quad_rg}(a) is the correct flow. 

{\em Hubbard Model:} Although our RG arguments apply independently of the
form of
short-range interactions~\cite{bernal2016ph:fn}, to provide a concrete lattice realization of
our predicted emergence of the Dirac phase from the QBT we present a detailed
numerical study of this phenomena in the Hubbard model on the Bernal stacked
honeycomb bilayer. Defining the electron creation operator on
sublattice A(B), site $i$, spin $\sigma$ and layer $\ell$ as $a^\dagger_{i\sigma \ell}$($b^\dagger_{i\sigma \ell}$), the Hamiltonian takes the form~\cite{mccann2013:bilrev},
\begin{eqnarray}
\label{eq:Hbsb}
   H_{\rm BSB} & = & 
     - t\sum_{\langle ij \rangle {\ell} }  a^\dagger_{i\sigma \ell} b^{\phantom{\dagger}}_{j\sigma \ell}
     - t_\perp\sum_{i} a^\dagger_{i \sigma 2} b^{\phantom{\dagger}}_{i
                     \sigma 1}  + {\rm h.c.} \nonumber\\
   & + & U \sum_{i,\ell} \left(
       a^\dagger_{i\uparrow \ell} a^{\phantom{\dagger}}_{i\uparrow \ell}
       a^\dagger_{i\downarrow \ell} a^{\phantom{\dagger}}_{i\downarrow \ell} 
     + b^\dagger_{i\uparrow \ell} b^{\phantom{\dagger}}_{i\uparrow \ell} 
       b^\dagger_{i\downarrow \ell} b^{\phantom{\dagger}}_{i\downarrow \ell} \right ),
\end{eqnarray}
where $\langle ij \rangle$ are bipartite nearest neighbors on the honeycomb
layers.  We choose $t=t_\perp=1$, fixed throughout the manuscript (results for ${t/t_\perp \neq 1}$ are presented in the SM). This model has been studied previously where it was
found that for $U/t>3$ the model has N\'eel order. For $U/t<3$ although
no numerical evidence was found for the presence of N\'eel order in
extrapolations of the order parameter, it
was assumed based on RG calculations that the model was
weakly N\'eel ordered~\cite{lang2012:bil}, supporting the RG flow in
Fig.~\ref{fig:quad_rg}(a) on the $\alpha=0$ axis. We have already
presented arguments that invalidate this conclusion and suggest that
the correct RG flow for the BSB is Fig.~\ref{fig:quad_rg}(b).  To
further back up our RG argument, we now present clear numerical
evidence that contrary to the conclusion of \cite{lang2012:bil},    $H_{\rm BSB}$  has a finite
coupling phase transition at which magnetism appears~\cite{bernal2016ph:fn2}. 

\begin{figure}[!t]
\centerline{\includegraphics[angle=0,width=\columnwidth]{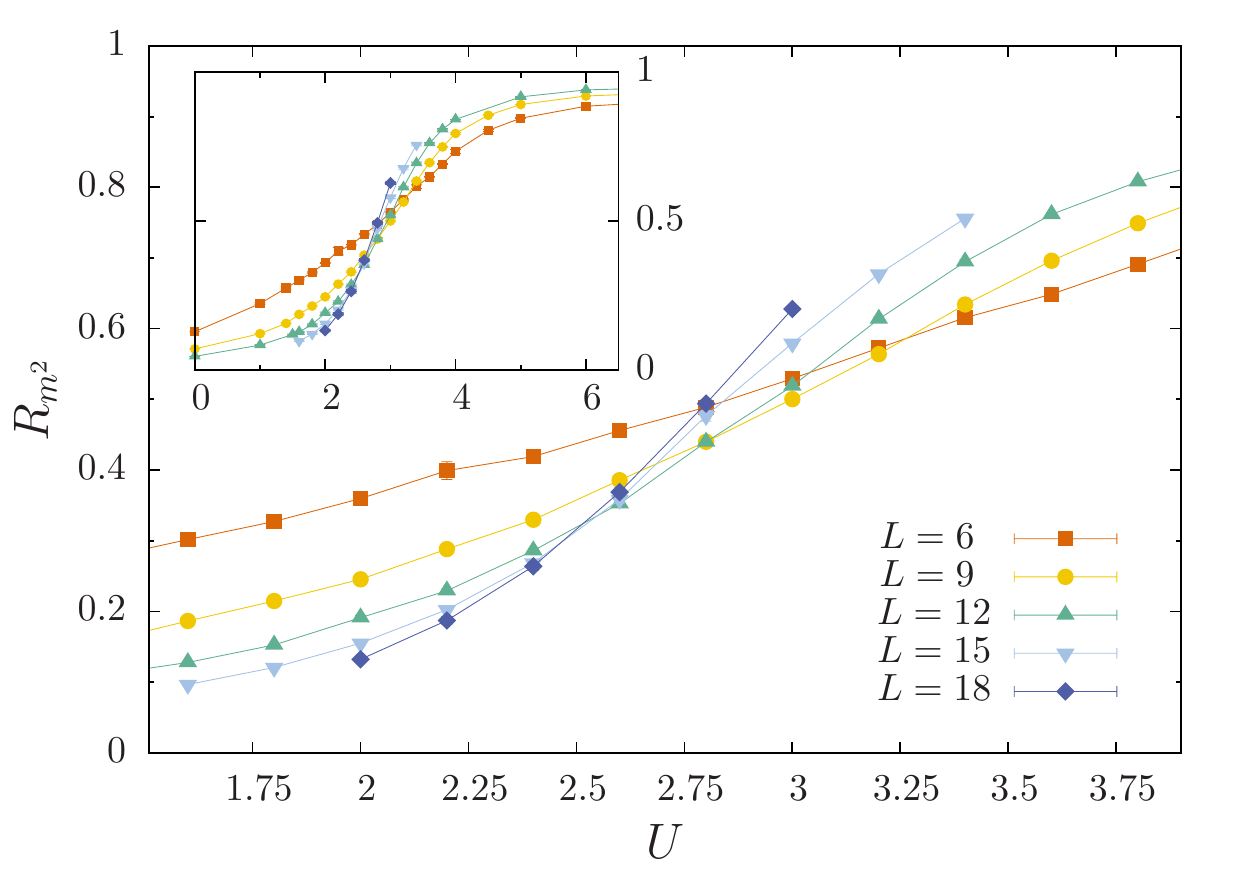}}
\caption{Correlation ratio $R_{ m^2}$, Eq.~(\ref{eq:Rm2}), close to the phase
  transition. The inset has a broad
  range of $U$ showing how $R_{m^2}$ reaches its asymptotes of 0 in
  the non-magnetic phase and 1 in the magnetic phase.  At a
  continuous transition $R_{m^2}$ is expected to be volume
  independent and cross at a universal value. The main panel show a zoom-in close to the critical
  point. The crossing point is at $U_c\approx 2.5$
  with very small drifts on the largest systems sizes.}
\label{fig:Rcross}
\end{figure}

To connect $H_{\rm BSB}$ with the long-wavelength description, Eq.~(\ref{eq:lw}), with only
$t$, $t_\perp$ hopping and $U=0$, the system realizes Eq.~(\ref{eq:lw})
with $\alpha=0$, a QBT arises described by the fixed point Q. Crucially, as noted earlier there is no symmetry that protects
$\alpha=0$. Indeed, trigonal hopping of the form $-t_{3} \sum_{\langle ij
  \rangle\sigma}b^\dagger_{i \sigma 2} a^{\phantom{\dagger}}_{j \sigma 1}$ preserves 
the symmetries of the bilayer and nonetheless has an
$\alpha$ term, which makes the dispersion at low energies linear~\cite{mccann2006:bil}. 

We emphasize that all our simulations are carried out
with $H_{\rm BSB}$ itself and with no $t_3$
hopping. Thus $H_{\rm BSB}$  can be thought of as lying along the
$\alpha=0$ axis in Fig.~\ref{fig:quad_rg}(a,b). The two RG
scenarios (a) and (b) make strikingly different prediction for the
phase diagram of $H_{\rm BSB}$. In
scenario (a) $H_{\rm BSB}$ should flow to strong coupling even for
arbitrarily small-$U$ causing the system to N\'eel order for all
$U$. In scenario (b), N\'eel order turns on at a
finite order-one coupling through a relativistic quantum critical
point. We now provide evidence using unbiased QMC
simulations (the method used is reviewed in~\cite{Assaad08:lnp}) on $H_{\rm BSB}$  that as argued above RG scenario (b) is realized.

For ${U/t \gg 1}$ we obtain the ${S=1/2}$ Heisenberg model on the BSB, which has  N\'eel order~\cite{lang2012:bil}. We focus on the fate of the N\'eel
order as $U$ is weakened. 
In Fig.~\ref{fig:msq}, we present data for the spin structure factor,
${S_{\rm AFM} (\mathbf{k}) \equiv
  \sum_{\mathbf{r}}\E^{\I\mathbf{k}\cdot \mathbf{r}} }\langle {\bf \cal{S}}
    ({\bf r} )\cdot {\bf\cal{ S}} ({\bf 0})\rangle$, where ${\bf
      \cal{S}} ({\bf r} )=\sum_{\ell}{\left(\mathbf{S}_{\ell
          A}({\bf r} )-\mathbf{S}_{\ell B} ({\bf r} )\right)}$ is the
    unit cell anti-ferromagnetic order parameter, 
for various values of $U$. We expect a N\'eel ordered state to
have a Bragg peak in the spin structure factor at the ${\bf \Gamma}$ point. As shown in
Fig.~\ref{fig:msq}(b) for ${U=3}$,
this is indeed the case. We find in Fig.~\ref{fig:msq}(c) that
$m^2\equiv S_{\rm AFM}({\bf \Gamma})/L^2$ versus $1/L$ extrapolates to a finite value,
indicating the divergence of the Bragg peak height as expected. As
shown in (b), at ${U=2}$ the peak gets rounded
out. Consistent with this rounding out, we find in Fig.~\ref{fig:msq}(c)
that $m^2$ appears to extrapolate to zero in
the thermodynamic limit for small $U$, indicative of a phase transition.

\begin{figure}[!t]
\centerline{\includegraphics[angle=0,width=\columnwidth]{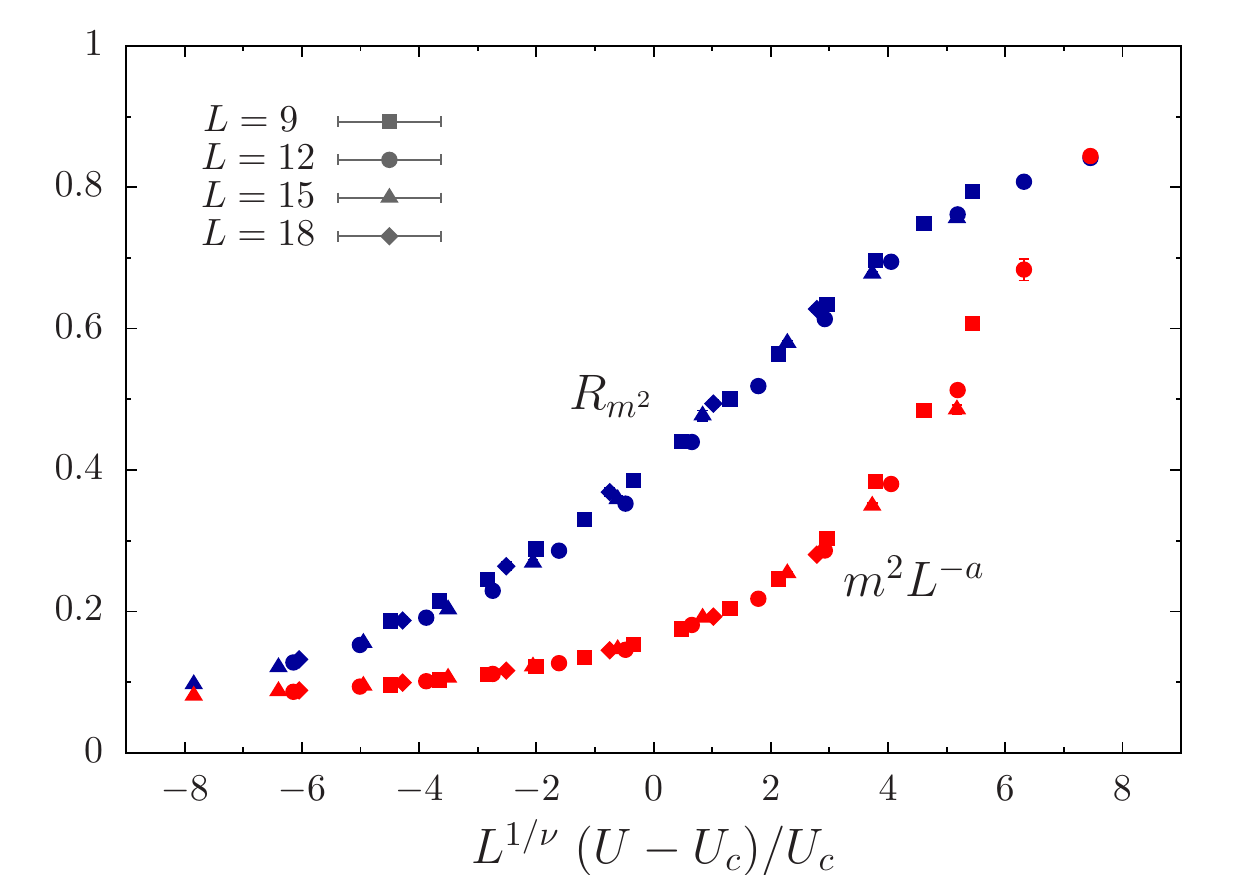}}
\caption{Collapse of the magnetic data, $m^2$ and $R_{m^2}$ close to the
  critical point. Parameters used for the collapse are $U_c=2.6$,
  $\nu=0.9$ and $a=0.3$.}
\label{fig:magclps}
\end{figure}
 
To study the long-distance ordering quantitatively rather than
extrapolate $m^2$ (which is notoriously hard when $m^2$ is small~\cite{toldin2015:fss}), we
study the correlation ratio $R_{m^2}$ of the structure factor at the
ordering momentum (the Bragg peak) and the momentum closest to it
$\mathbf{b}/L$ ($\mathbf{b}$ is the shortest reciprocal lattice vector )
\begin{equation}
\label{eq:Rm2}
R_{m^2} = 1- \frac{S_{\rm AFM}({\bf \Gamma + b}/L)}{S_{\rm AFM}({\bf \Gamma)} }\;.
\end{equation}
This quantity scales to 1 in an ordered phase, scales to 0 in a
phase disordered phase, and crosses at a universal value for different $L$ at a quantum critical point, providing a sensitive
numerical test for magnetic ordering~\cite{kaul2015:son}.
The data in Fig.~\ref{fig:Rcross} shows a crossing at a value
of $U_c\approx 2.5$ and does not drift significantly on all but the
smallest lattices, providing strong evidence for a
finite-coupling phase transition (see SM for more details).

\begin{figure}[!t]
\centerline{\includegraphics[angle=0,width=\columnwidth]{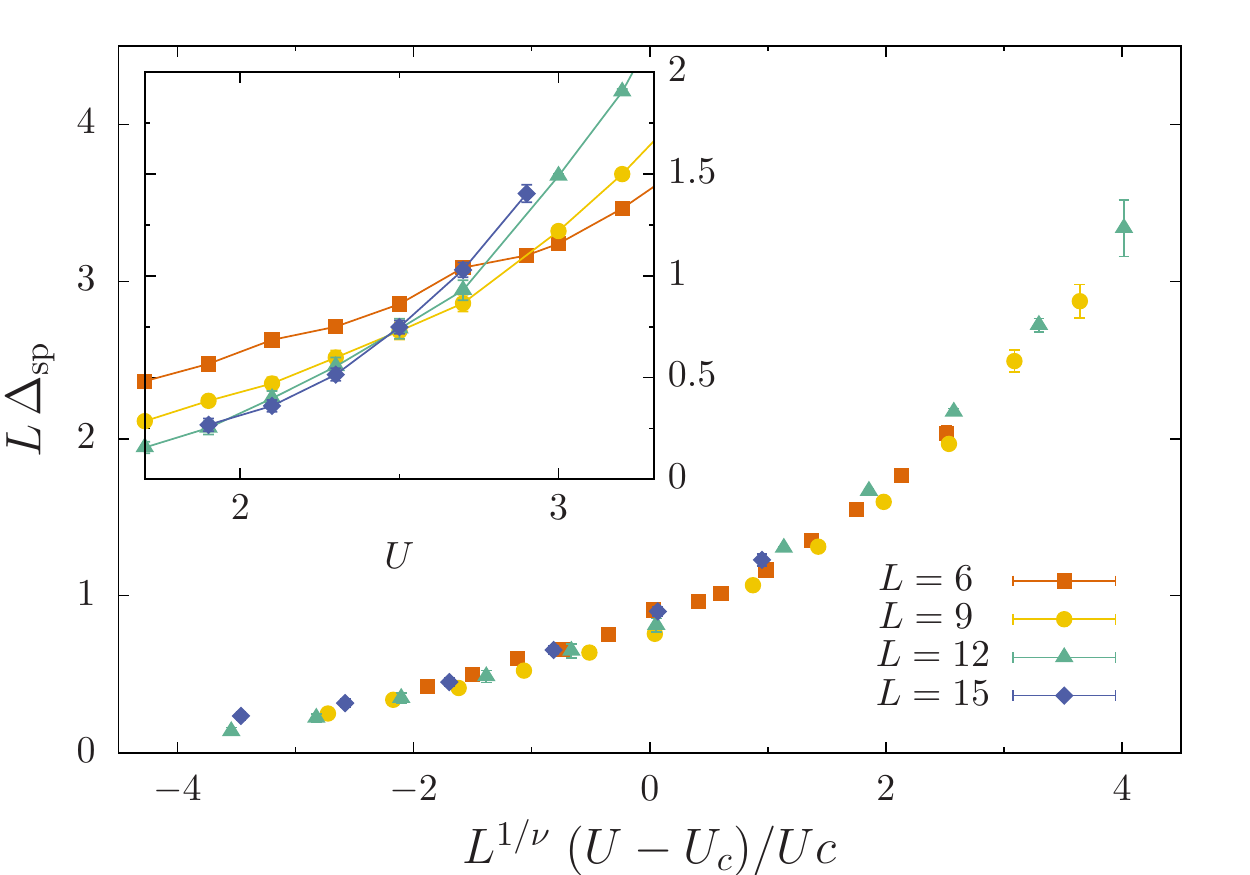}}
\caption{The collapse of the single-particle gap $L\Delta_{\rm sp}$ close to
sthe phase transition with the same critical parameters as in
Fig.~\ref{fig:magclps} and $z=1$. The inset shows the single-particle gap
$L\Delta_{\rm sp}$ as a function of $U$ which  exhibits a crossing at a coupling of $U_c\approx
2.5$, consistent with the
expected scaling form with $z=1$.  }
\label{fig:gap}
\end{figure}

Field theoretic arguments suggest that the transition between the Dirac
phase and the N\'eel phase should be in the
GN universality class. The key hallmarks of this kind of transition are: the gap in the Dirac fermions and the onset of magnetism take place
at the same value of $U$ and the emergence of Lorentz invariance or
$z=1$ scaling. To test for
these features we study the collapse of the magnetic data ($m^2$ and
$R_{m^2}$) and single particle gap $\Delta_{\rm sp}$, close to the
transition. We extract $\Delta_{\rm sp}$ from the decay of
the imaginary time displaced Green function using standard methods
\cite{Feldbacher01,Lang14}. We collapse the magnetic data in Fig.~\ref{fig:magclps}, with the
scaling forms 
$m^2 = L^a {\cal F}_{{m^2}} \left[L^{1/\nu}(U-U_c)/U_c\right]$ and 
$R_{m^2} = {\cal F}_{R_{m^2}} \left[L^{1/\nu}(U-U_c)/U_c\right]$,
since the $R_{m^2}$ has no
scaling dimension. Corresponding windows for the parameters that give acceptable collapses are $U_c=2.6(1)$,
  $\nu=0.9(2)$ and $a=0.3(1)$.
For the gap, using standard finite-size scaling Ansatz, we expect
${\Delta_{\rm sp} = L^{-z} { \cal  F}_\Delta \left[L^{1/\nu}(U-U_c)/U_c\right]}$. We have verified
that our data is consistent with this Ansatz with the values of the
parameters extracted from the collapse for $R_{m^2}$ and
$m^2$, and $z=1$ in Fig.~\ref{fig:gap}. Independently, we obtain acceptable
collapses for $\Delta_{\rm sp}$ for the ranges: $U_c=2.5(2)$,
  $\nu=1.0(2)$ and $z=0.9(2)$. The inset in Fig.~\ref{fig:gap}
shows our data for $L\Delta_{\rm sp}$ versus $U$ which should cross at
the critical point if $z=1$. We see a crossing in our data which is
consistent with the crossings of $R_{m^2}$ indicating the existence
of a single transition at which both the magnetization turns on and
the Dirac fermions get a mass. This is compelling evidence that the transition has $z=1$, in contrast
to the $z=2$ scaling at the Q fixed point. Taken together, the
identification of the quantum critical point with the GN class provides further evidence that $H_{\rm BSB}$
realizes the RG flow shown in Fig.~\ref{fig:quad_rg}(b). 

We emphasize that our RG flow,
Fig.~\ref{fig:quad_rg}(b) is only schematic: The point marked as
D is used to describe collectively fixed points with
a linear dispersion but which differ in the number and
location of the Dirac cones. Even though we have provided
strong evidence for a linear dispersion, we are unable to
resolve from our QMC simulations which of these fixed
points is realized in $H_{\rm BSB}$. Our focus here has been on the phase
diagram when $\alpha=0$ in the microscopic model, i.e. the free
problem has a QBT, such as in $H_{\rm BSB}$. The evolution of the phase diagram with an explicit
trigonal warping term in $H_{\rm BSB}$ and a more detailed study of
the electronic structure will be presented separately.
 
Our results have  important implications for the interpretation of experiments on bilayer
graphene.
Short range interactions are the correct model for the experiments
in the presence of screening. In this situation our finding implies
that symmetry breaking is not an immediate consequence of interactions,
contrary to what has generally been assumed. Indeed bilayer graphene responds to interactions in much the
same way as the single layer: symmetry breaking will
only set in at a finite coupling strength. Making the widely accepted
assumptions that interactions are weak and trigonal warping is absent
or even weaker, our results predict that that bilayer graphene with
screened interactions 
will be in a Dirac phase without any symmetry breaking.  We emphasize however that while our
RG argument applies
independently of the form of the short-range
interactions, it does not go through in the
presence of long-range Coulomb interactions, which by power counting
are relevant at the QBT fixed point and hence can plausibly cause a weak
coupling instability. Interestingly, this last fact implies that
the symmetry breaking that has been detected in experiment on suspended bilayer
graphene is likely a consequence of the long-range tail of the Coulomb interactions~\cite{feldman2009:bilnat,martin2010:bil,weitz2010:science}.
 
We thank F.~Assaad, C.~L.~Kane, C.~Honerkamp, A.~C.~Potter, O.~Vafek,
S.~Wessel, A.~Vishwanath and K.~Yang for useful
discussions. We acknowledge NSF DMR-1056536 (SP and RKK), NSF
DMR-1306897 (GM), US-Israel BSF 2012120 (GM) for financial support, and
NSF XSEDE DMR-150037, SuperMUC at Leibniz Supercomputing Centre and
JURECA at J\"ulich Supercomputing Centre (JSC) for generous computer allocations.

\putbib[fermion.bib]

 \end{bibunit}

%\bibliography{fermion}

\begin{thebibliography}{25}
\bibitem[{ber({\natexlab{a}})}]{bernal2016ph:fn}
\bibinfo{note}{Our RG argument assumes the model under study has particle-hole
  symmetry. This symmetry is present in all basic models of graphene systems
  including the $H_{\rm BSB}$ studied here, Eq.~(\ref{eq:Hbsb}).}
\end{thebibliography}

\clearpage

\begin{bibunit}[apsrev]

\begin{widetext}
\begin{center}
{\large \bf SUPPLEMENTAL MATERIALS } \\
\vspace{5mm}
{\bf ``Interaction induced Dirac fermions
  from quadratic band touching in bilayer graphene''}\\
\vspace{2mm}
 Sumiran Pujari,$^1$
Thomas C. Lang,$^2$ Ganpathy Murthy$^1$ and Ribhu K. Kaul$^1$\\
\vspace{2mm}
{\it $^1$Department of Physics \& Astronomy, University of
  Kentucky, Lexington, KY-40506-0055\\
$^2$Institute for Theoretical Physics, University of Innsbruck, 6020 Innsbruck, Austria}
\vspace{5mm}
\end{center}
\end{widetext}

\begin{figure}[h]
\includegraphics[trim=0mm 100mm 0mm 0mm, clip=true, 
width=\linewidth]{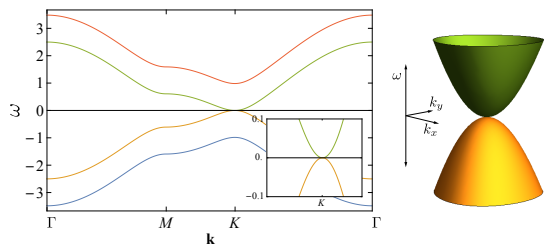}
\caption{ \label{fig:dsp} Electronic dispersion $\omega ({\bf k})$  for Bernal stacked
  bilayer graphene, corresponding to the $U=0$ limit of
  Eq.~(\ref{eq:Hbsb}) with $t=t_\perp=1$. The zoom-in figure on the
  right shows the quadratic band touching of the low energy bands at
  the {\bf K} point. More details can be found in~\cite{mccann2006:bil}. }
\end{figure}

\section{RG Arguments}

\subsection{Generation of the linear kinetic term in perturbation theory}
\label{supp1}

In this section we will show that even if one starts with a pure
quadratic term at $\bK$ and $\bK'$, second-order perturbation theory
in the Hubbard interaction generates a linear term. First we collect a
few pertinent facts. We will focus on momenta near one of the
quadratic points, say $\bK$ for specificity. The bare kinetic term at
$\bK+\bk$ looks like
\beq
   \left(\!
      \begin{array}{cc} 
         0 & -f^*(\bk)^2\\
         -f(\bk)^2&0
     \end{array}
   \!\right)\;.
\eeq
For $|\bk|$ small we can expand $f(\bk)$ as 
\beq
   f(\bk)=-\frac{\sqrt{3}}{2}(k_x- \I k_y)+\frac{1}{8}(k_x +\I k_y)^2 \;,
\eeq 
and hence we can approximate 
\beq 
   \big[f(\bk)\big]^2=C_0k_-^2+C_1\bk^2 k_+\;,
\eeq
where $k_{\pm}=k_x\pm \I k_y$. We will rescale momenta to make the
number $C_0=1$ for future convenience.

In the Euclidean path integral the bare Green's function is
\beq
   \cG_{ij}(\omega,\bk)
   = \frac{\delta_{ij}}{\omega^2+|f(\bk)|^4}
   \left(\!
    \begin{array}{cc}
       \I\omega& -f^*(\bk)^2\\
       -f(\bk)^2&\I\omega
   \end{array}
   \!\right)\;.
\eeq
Here $i,j=1,\dots,N$ label the flavours. We will confine ourselves to the momentum independent interactions
which are marginal in the bare theory. As categorized by Vafek (Ref. 8 of the main text), there
are many classes of such interactions consistent with the
symmetries. We will take the simplest one, the Hubbard interaction
with $N$ flavors, with comes with the unit matrix. Once again
confining ourselves only to fields $\psi_{ia}$ at the $\bK$ point, with
$i=1,\dots,N$ labelling the flavors and $a=1,2$ labelling the
two-dimensional space of bands touching at the $\bK$ point we have the
vertex
\begin{multline}
   U\int d\bk_1\, d\bk_2\, d\bq \sum\limits_{i,j,a,b}\\ 
   : {\bar \psi}_{ia}(\bk_1-\bq)\, \psi_{ia}(\bk_1)\, {\bar \psi}_{jb}(\bk_2+\bq)\, \psi_{jb}(\bk_2):\;.
\end{multline}
The two diagrams contributing to the self-energy ${\Sigma(\mathbf{k})}$ to
first order in $U$ (1-loop) are
\begin{figure}[!h]
   \center\includegraphics[width=0.67\columnwidth]{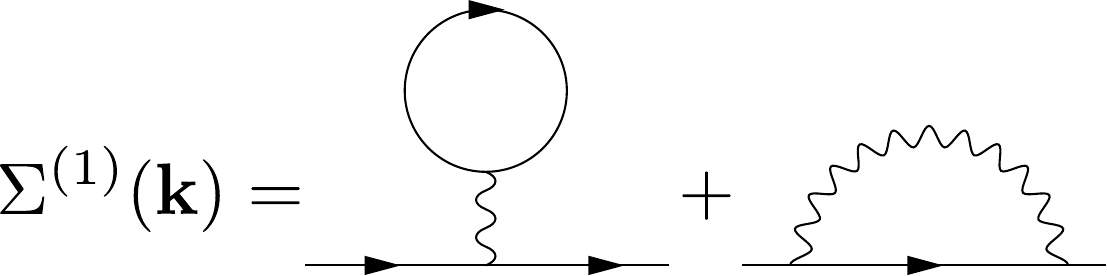}\;.
\end{figure}\\
It is easily seen that both diagrams give momentum-independent corrections, and hence cannot generate a linear
term in $\bk$ near the $\bK$ point.
The self-energy diagrams to second order in $U$ are
\begin{figure}[!h]
   \center\includegraphics[width=0.96\columnwidth]{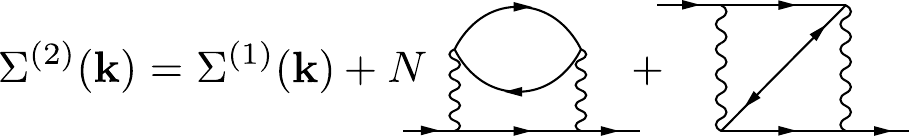}\;.
\end{figure}\\

It is clear that they have different dependences on $N$ and hence
cannot cancel for generic $N$. Therefore it is sufficient to examine
the first one, setting the external frequency to zero, since we want
the dependence on the spatial momentum. This diagram has the
expression
\begin{multline}
   \Sigma_{ij}(\bk) = \frac{1}{(2\pi)^6} \int d\omega\, d\Omega\,d^2p\,d^2q\, \cG_{ij}(-\Omega,\bk-\bq)\\
   \mathrm{Tr}\left[\cG\!\left(\omega+\frac{\Omega}{2},\bp+\frac{\bq}{2}\right)
   \cG\!\left(\omega-\frac{\Omega}{2},\bp-\frac{\bq}{2}\right)\right]\;. 
\end{multline}
Of course the spatial momentum integrals must be cut off in some
rotationally invariant way, by including factors such as
$\exp{(-\bk^2/\Lambda^2)}$, say. The crucial observation is that the result of carrying out the $\bp$ and $\omega$ integrals is a function which is rotationally invariant in $\bq$. So
\begin{multline}
   \frac{1}{(2\pi)^6} \int d\omega\, d^2p\,\mathrm{Tr}\left[\cG\!\left(\omega+\frac{\Omega}{2},\bp+\frac{\bq}{2}\right)\right.\\
   \left.\cG\!\left(\omega-\frac{\Omega}{2},\bp-\frac{\bq}{2}\right)\right] = N\ F(\Omega,\bq^2)\;.
\end{multline}

Now let us concentrate on the off-diagonal (in the $a$ index) part of
the self-energy, since that is where the linear term will manifest
itself.
\beq
   \big[\Sigma_{ij}(\bk)\big]_{12}=-\frac{N U^2}{(2\pi)^3} \int d\Omega\,d^2q\, F(\Omega,\bq^2) \frac{\big(f^*(\bk-\bq)\big)^2}{\Omega^2+|f(\bp-\bq)|^4}\;.
\eeq
We focus on very small $\bk$ close to the $\bK$ point, expanding the
integrand in powers of $\bk$, keeping only up to first order in $\bk$.
Recalling that we have scaled momenta such that $C_0=1$ and up to higher orders we find the numerator to be 
\beq
   q_+^2-2k_+q_+-C_1\bq^2q_-+C_1(\bq^2k_-+2\bk\cdot\bq q_-)\;,
\eeq
while the denominator is
\begin{multline}
   \Omega^2+(\bq^2)^2-4\bq^2(\bk\cdot\bq) \\
   +C_1\left[3\bq^2(k_+q_+^2+k_-q_-^2)+2\bk\cdot\bq(q_+^3+q_-^3)\right]\;.
\end{multline}
We can now carry out the angular integral in $d^2q$. It is clear that
no terms linear in $\bk$ survive the angular integral in the expansion
of the denominator. However, a linear term does survive to first order in $\bk$ in the numerator. This is 
\beq
   -\frac{2NC_1U^2}{(2\pi)^3}\ k_-\int d\Omega\,d^2q\, F(\Omega,\bq^2)\, \frac{\bq^2}{\Omega^2+(\bq^2)^2} \;.
\eeq
We have demonstrated that a term linear in $\bk$ is generically present in the two-loop correction to the self-energy.

\subsection{Cartoon RG flow including the generation of the linear term at the QBT}
\label{cartoonRG}

We now examine what a cartoon of the RG flow might look like when we
take the two-loop interaction generated linear term into account.
Let us consider a single 4-point coupling $g$ which is marginally
relevant at the bare QBT fixed point with only quadratic terms in the
kinetic energy. We will call the linear term in the kinetic energy
$\alpha$. The coupled RG equations will have the generic form
\begin{equation}
	\frac{dg}{d\ell}=g^2 \;,\quad\quad
	\frac{d\alpha}{d\ell}=\alpha+Ag^2\;,
\end{equation}
where $A$ is some constant. The second equation follows from the fact
that a linear term is relevant with scaling dimension 1 in the bare theory with a pure quadratic kinetic term. The initial conditions are
$\alpha(0)=0;\ g(0)=g_0\ll 1$. Solving the first equation we find
\beq
   g(\ell)=\frac{g_0}{1-g_0\ell}\;,
\eeq
so the scale at which $g(\ell_g)\approx 1$ is $\ell_g\approx\frac{1}{g_0}$. On the other hand we can solve the second equation with the initial condition as 
\beq
   \alpha(\ell)=A e^{\ell}\int\limits_{0}^{\ell} d\ell'\, g^2(\ell')\, e^{-\ell'}\;.
\eeq
Clearly 
\beq
   \alpha(\ell)> A\,g_0^2\,e^{\ell}\int\limits_{0}^{\ell} d\ell'\, e^{-\ell'} = A\,g_0^2\,(e^{\ell}-1)\;.
\eeq
The scale at which $\alpha(\ell)\approx 1$ is therefore
$\ell_{\alpha}\approx \log{Cg_0^2}$. 
For small enough $g_0$ it is clear that $\ell_{\alpha}$ is
parametrically smaller than $\ell_g$, which means that the linear term
becomes important far \textit{earlier} in the RG flow than $g$ as one flows to the
infrared. We interpret this as a flow to a fixed point with a linear
dispersion where interactions are well known to be irrelevant.

%\subsection{Fine-tuning and reentrant transitions}
%
%As seen in Fig 1b in the text, for generic initial conditions for
%$g_0,\ \alpha_0$ close to the origin the system flows to the Dirac
%fixed point where interactions are irrelevant. However, it is possible
%to fine-tune the system to lie in the narrow wedge between the two red
%lines in Fig 1b. In this region, an arbitrarily weak initial value of
%$g_0$ will flow to strong coupling, in line with the results of Vafek. 
%
%Given the validity of Fig 1b, two interesting phenomena should be
%observed.  First, suppose one keeps the Hubard $U$ fixed at a value
%smaller than the critical value $U_c$ and scans in $\alpha$. For a
%narrow range of negative $\alpha$ symmetry-breaking should be
%observed. Second, suppose one keeps $\alpha$ small, negative and fixed
%and scans in $U$. Now, for extremely tiny $U$ the system flows to the
%$\alpha<0$ Dirac fixed point. For a narrow intermediate range
%$U_{c1}<U<U_{c2}$ the initial conditions fall within the two red
%separatrices, and symmetry-breaking should be observed. For a broader
%range $U_{c2}<U<U_{c3}$ symmetry should be restored with the system flowing to the $\alpha>0$ Dirac fixed point, and for
%$U>U_{c3}$ symmetry should be broken.

\section{Additional unbiased QMC Data}

\subsection{Phase diagram in $t_p$-$U$ plane}

\begin{figure}[h]
\includegraphics[trim=0mm 0mm 0cm 0mm, clip=true, 
width=0.9\linewidth]{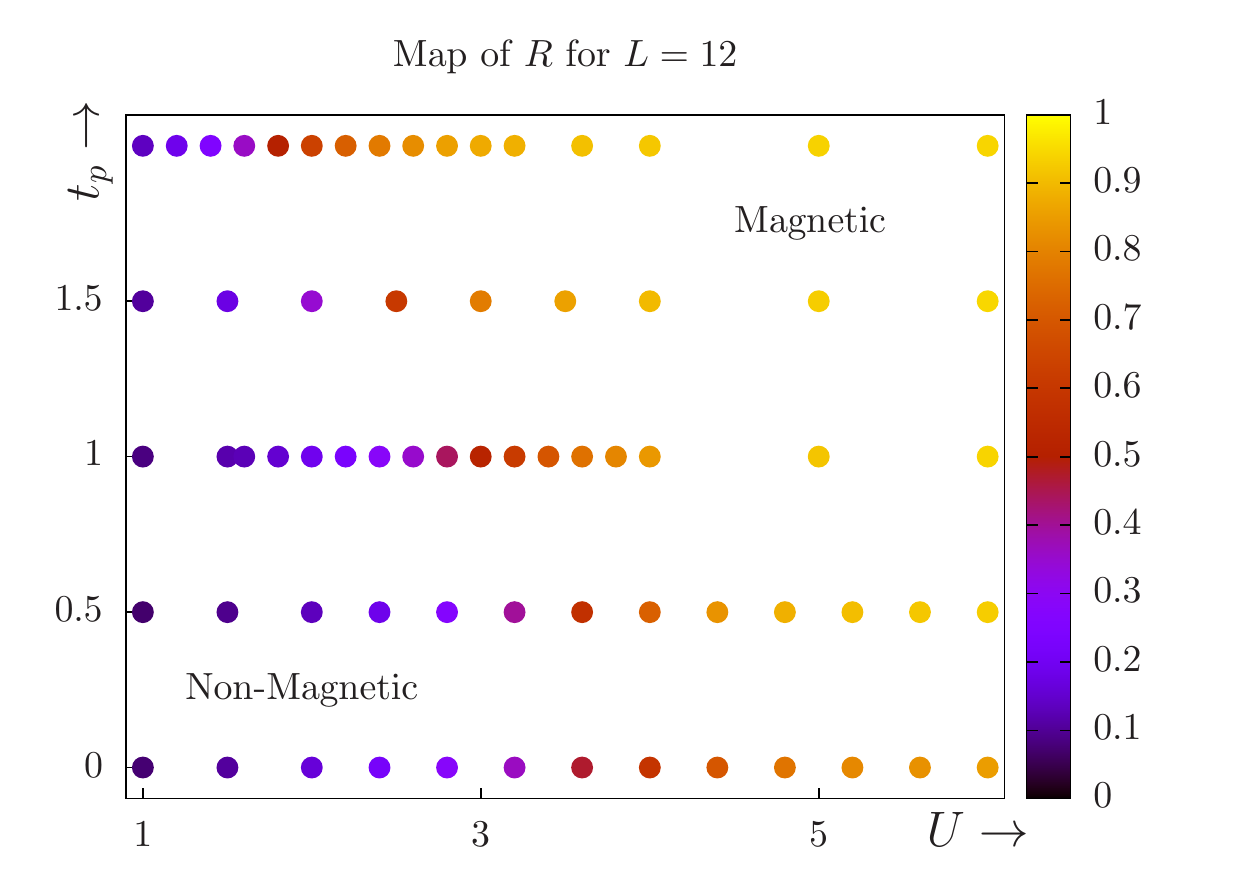}
\caption{ \label{fig:Rmap}
A map of correlation ratio $R_{m^2}$ vs $t_p$ and $U$ in units
of $t=1.$ for $L=12$.
}
\end{figure}

In Fig. \ref{fig:Rmap} we show a map of the correlation ratio
$R_{m^2}$ (defined in the main text) for $L=12$ as $t_p/t$ is tuned.
For the case of $t_p/t=1.$ which was the focus in the main text,
the critical value of $U$ was found to be $U_c \approx 2.6$.
Around this interaction strength
the correlation ratio $R$ is approximately
$R_c \approx 0.4$ as can be seen in the above figure. By tracking where $R_{m^2}$
crosses this approximate $R_c$ we can see how the transition in the bilayer
smoothly evolves in to that of the monolayer. Since we argue for Gross-Neveu
criticality for the bilayer same as the monolayer
in the main text, we may expect this smooth evolution of the critical
point. Of course it is possible that there are Lifshitz like
transitions where the number of Dirac nodes change inside the ``non-magnetic'' phase.

%But we should keep in mind that the weak-coupling phase of bilayer is 
%physically different
%than that of the monolayer. The bilayer has four Dirac cones 
%per valley and the separation
%between can be tuned by tuning $t_p$ and $U$. The monolayer 
%has one Dirac cone per valley. Thus critical quantities like correlation
%exponent $\nu$ and anomalous exponent associated with the AFM order parameter
%$a$ need not \emph{a priori} be the same for the bilayer and the monolayer.
%This investigation is left to the future.

As $t_p/t \rightarrow \infty$, the quadratically touching bands 
become completely flat and we may expect that an infinitesimal
value of 
Hubbard interaction strength immediately 
opens a gap. The evolution of $U_c$ as $t_p/t$ increases in 
in agreement with this expectation for the values of $t_p/t$ studied.

\subsection{Correlation ratio for Square lattice Hubbard model at half-filling}

\begin{figure}[h]
\includegraphics[trim=0mm 0mm 0cm 0mm, clip=true, 
width=0.9\linewidth]{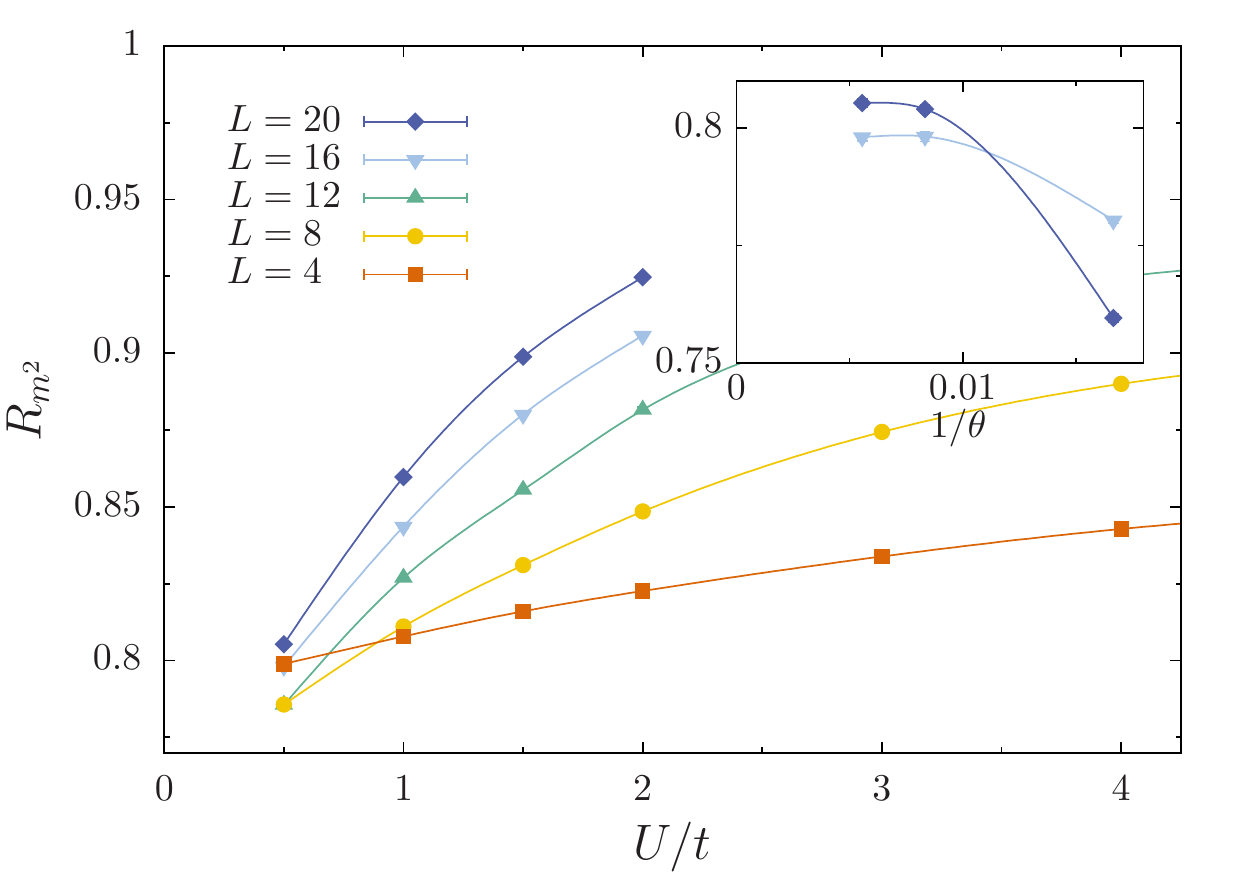}
\caption{ \label{fig:binder_square}
 Plot of $R_{m^2}$ vs. $U/t$ for half-filled Hubbard model
on the square lattice. 
Note that contrary to the Honeycomb
Bilayer case, the $R$ value are greater than 0.75 for all values of $U/t$ studied.
Furthermore they are increasing with
increasing system sizes and increasing projection lengths $\theta$ (see
inset).}
\end{figure}

In order to contrast the finite-$U$ transition to the antiferromagnetic insulator
on the honeycomb bilayer discussed in the main text, as evidenced by the
correlation ratio $R_{m^2}$, we look at the corresponding quantity in the SU(2) Hubbard model at half-filling on the square-lattice.
Here we have a nested Fermi surface instead
of Fermi points. As shown in Fig. \ref{fig:binder_square}, we see that there
is antiferromagnetic ordering for all values of interaction $U$ 
and system sizes $L$ studied. This data is consistent with a 
 $U \rightarrow 0^+$ instability. The results for $R_{m^2}$ at the smallest value of $U=0.5$ still increase with growing system sizes such that there is significant drift of the crossing points towards zero. This behavior is in contrast to the situation on the bilayer, where $R_{m^2}$ clearly decreases as a function of system sizes for ${U<2.5}$. Furthermore note the large values of $R_{m^2}$ close to 1, far from the magnetically disordered limit of 0.

\subsection{Comparison with Exact Diagonalization}
In Table \ref{tab:EDvsQMC1},\ref{tab:EDvsQMC2},\ref{tab:EDvsQMC3} we tabulate several 
observables computed using a
ground state projection version of determinantal QMC 
with
Exact Diagonalization (ED) for several observables. The system size 
is $L_x=2$, $L_y=1$ which corresponds to 8 sites. The two unit cells 
are indexed as $\mathbf{r_1}=(0,0)$ and $\mathbf{r_2}=(1,0)$.
The 4 lattice sites
per unit cell are further indexed by an integer $\mu$ ranging
from 1 to 4. 
 Parameters of the 
Hamiltonian were chosen to be $N=2$ and $t=t_p=U=1.0$.
$\theta$ represents the length of ground-state
projection $e^{-\theta H}$
applied on a trial state to reach the ground state.
$d\tau$ represents the discretization of projection length
and gives rise to Suzuki-Trotter errors as seen in
Table \ref{tab:EDvsQMC1},\ref{tab:EDvsQMC2},\ref{tab:EDvsQMC3}.

\begin{table}[h!]
\centering
\begin{tabular}{ |c|c|c|c|c| }
\hline \hline
\: \: \:  Observable \: \: \:  & \: \: \: ED \: \: \: & \: \: \: QMC \: \: \: & $\: \: \: \theta \: \: \:  $ & $\: \: \: d\tau\: \: \: $ \\ 
\hline \hline
 Total Energy & -14.7555... & -14.755(1) & 40 & 0.05    \\
  &  & -14.755(1) & 40 & 0.1    \\
 &  & -14.751(1) & 40 & 0.2    \\
 & & -14.755(1) & 80 & 0.05    \\
 & & -14.7553(6) & 80 & 0.1    \\
 & & -14.751(1) & 80 & 0.2    \\
 \hline
Kinetic Energy & 1.7648... & 1.7664(5) & 40 & 0.05    \\
 & & 1.7715(3) & 40 & 0.1    \\
 & & 1.7895(3) & 40 & 0.2    \\
 & & 1.7669(2) & 80 & 0.05    \\
 & & 1.7714(2) & 80 & 0.1    \\
 & & 1.7890(3) & 80 & 0.2    \\
 \hline
 Potential Energy & -16.5203... & -16.522(1) & 40 & 0.05    \\
 &  & -16.527(1) & 40 & 0.1    \\
 & & -16.5403(5) & 40 & 0.2    \\
 & & -16.5214(5) & 80 & 0.05    \\
 & & -16.5267(5) & 80 & 0.1    \\
 & & -16.5402(5) & 80 & 0.2    \\
   \hline \hline
\end{tabular}
\caption{Comparison between QMC and ED data
for Energy observables}
\label{tab:EDvsQMC1} 
\end{table}

 \begin{table}[h!]
\centering
\begin{tabular}{ |c|c|c|c|c| }
\hline \hline
\: \: \:  Observable \: \: \:  & \: \: \: ED \: \: \: & \: \: \: QMC \: \: \: & $\: \: \: \theta \: \: \:  $ & $\: \: \: d\tau\: \: \: $ \\ 
\hline \hline
  $\langle S_z(\mathbf{r_1},2) S_z(\mathbf{r_1},2)\rangle$ & 0.277468...
 & 0.27729(2) & 40 & 0.05 \\ 
  & 
 & 0.2766(1) & 40 & 0.1 \\ 
  & 
 & 0.27427(1) & 40 & 0.2 \\ 
  $\langle S_z(\mathbf{r_1},2) S_z(\mathbf{r_2},2)\rangle$ & 0.001907...
 & 0.001908(1) & 40 & 0.05 \\ 
  & 
 & 0.001881(3)  & 40 & 0.1 \\ 
  & 
 & 0.001843(2) & 40 & 0.2 \\ 
$\langle S_z(\mathbf{r_1},3) S_z(\mathbf{r_1},4)\rangle$ & -0.250728...
 & -0.25046(2) & 40 & 0.05 \\ 
  & 
 &  -0.24961(2)   & 40 & 0.1 \\ 
  & 
 & -0.24628(2) & 40 & 0.2 \\ 

 $\langle S_z(\mathbf{r_1},3) S_z(\mathbf{r_2},4)\rangle$ & -0.003163...
 & -0.00316(1)  & 40 & 0.05 \\ 
  & 
 &  -0.003141(4) & 40 & 0.1 \\ 
  & 
 & -0.003089(4) & 40 & 0.2 \\ 
  \hline \hline
\end{tabular}
\caption{Comparison between QMC and ED data
for (diagonal) Spin-spin correlations}
\label{tab:EDvsQMC2} 
\end{table}

  \begin{table}[h!]
\centering
\begin{tabular}{ |c|c|c|c|c| }
\hline \hline
\: \: \:  Observable \: \: \:  & \: \: \: ED \: \: \: & \: \: \: QMC \: \: \: & $\: \: \: \theta \: \: \:  $ & $\: \: \: d\tau\: \: \: $ \\ 
\hline \hline

  $\langle c^\dagger(\mathbf{r_1},1) c(\mathbf{r_1},2)\rangle$ & 0.467109...
 & 0.4671(1)  & 40 & 0.05 \\ 
  & 
 & 0.4670(1) & 40 & 0.1 \\ 
  & 
 & 0.46632(4) & 40 & 0.2 \\ 
 $\langle c^\dagger(\mathbf{r_1},1) c(\mathbf{r_2},2)\rangle$ & 0.023232...
 & 0.02322(5) & 40 & 0.05 \\ 
  & 
 & 0.02315(4) & 40 & 0.1 \\ 
  & 
 & 0.02282(3) & 40 & 0.2 \\ 
 $\langle c^\dagger(\mathbf{r_1},1) c(\mathbf{r_1},3)\rangle$ & -6.29608...E-9
 & 0.00003(3) & 40 & 0.05 \\ 
  & 
 & 0.00031(3) & 40 & 0.1 \\ 
  & 
 & 0.00255(2) & 40 & 0.2 \\ 
 $\langle c^\dagger(\mathbf{r_1},1) c(\mathbf{r_2},3)\rangle$ & 2.83007...E-9  
 & 0.00004(3) & 40 & 0.05 \\ 
  & 
 & 0.00025(3) & 40 & 0.1 \\ 
  & 
 & 0.00206(2) & 40 & 0.2 \\ 
  \hline \hline
\end{tabular}
\caption{Comparison between QMC and ED data
for (off-diagonal) single-particle Green's functions.}
\label{tab:EDvsQMC3} 
\end{table}

A projection length of $\theta=40$ is found to be sufficient.
Trotter errors are present for the spin correlations and Green's function
values, but for larger systems of interest (as shown in the next section) 
statistical errors dominate Trotter errors which is the regime we would
like to sit at in determinantal QMC computations.

\newpage

\subsection{Projection and Discretization Errors}

\begin{figure}[h!]
\includegraphics[trim=0mm 0mm 0cm 0mm, clip=true, 
width=0.45\linewidth]{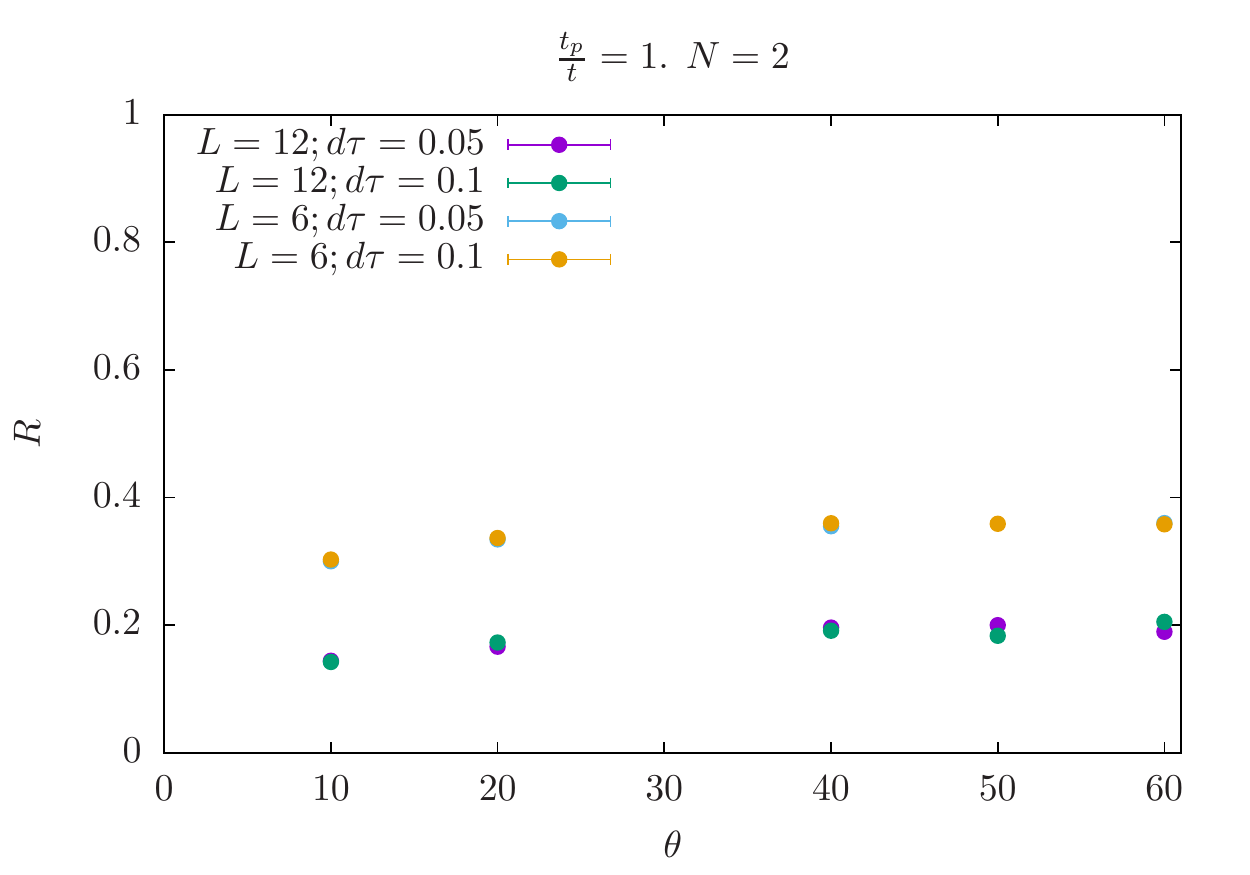}
\includegraphics[trim=0mm 0mm 0cm 0mm, clip=true, 
width=0.45\linewidth]{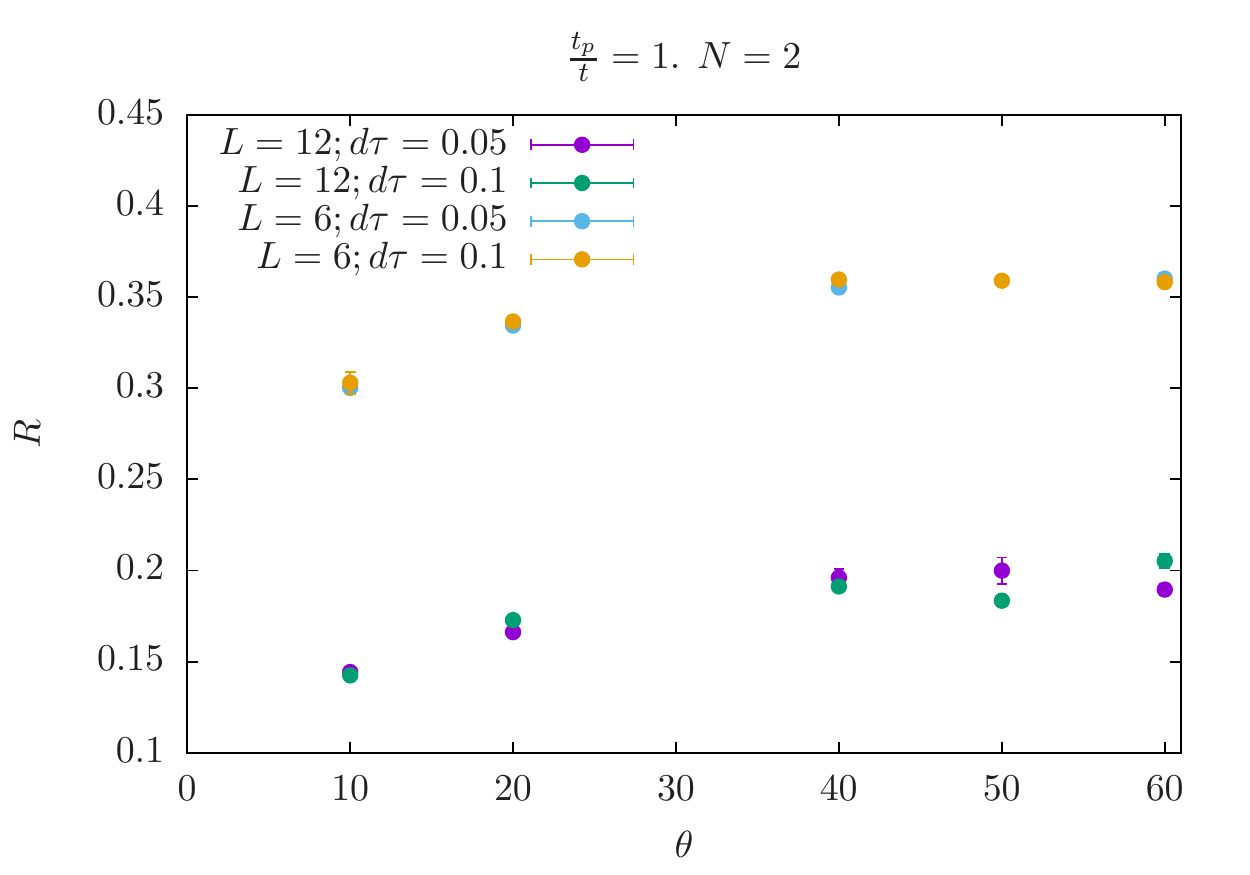}
\includegraphics[trim=0mm 0mm 0cm 0mm, clip=true, 
width=0.45\linewidth]{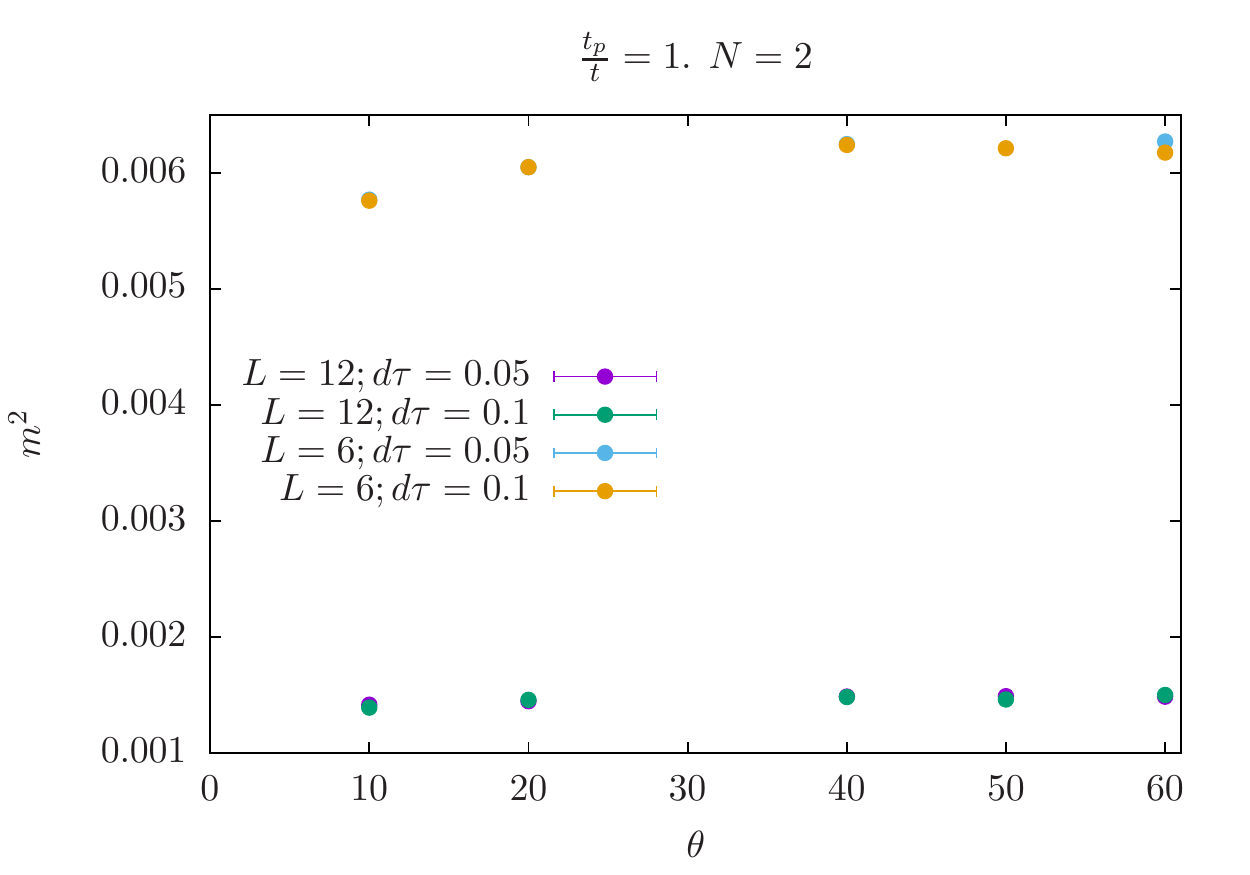}
\caption{ \label{fig:proj}
Projection length ($\theta$) dependence of correlation ratio $R$
(first two panels) and AFM order parameter $m^2$ 
for a representative value of Hubbard parameter ($U=2.$)
in the weak-coupling phase.
}
\end{figure}

In Fig. \ref{fig:proj}, we show the projection length ($\theta$) dependence
of the two main quantities of interest, namely the dimensionless
correlation ratio $R$ (first two panels)
and the antiferromagnetic order parameter
$m^2$ (last panel) as defined in the main text, 
for a representative value of Hubbard parameter ($U=2.$)
in the weak-coupling phase. In the strong coupling phase,
 projection to the ground states is even faster
because of finite single particle gap due to presence of (N\'eel)
order. 
For both the quantities, we see that statistical errors
completely dominate Trotter errors, since for each value of 
$\theta$ the data points corresponding to the two values of $d\tau$
are \emph{always} within QMC errorbars of each other. This is 
pragmatic from a QMC point
of view when statistical errors dominate systematic errors
for systems of interest to us.

For $R$ we see a $\theta$ dependence (second panel
of Fig. \ref{fig:proj})
which evens out for $\theta \geq 40$,
and all the data points for a given system size $L$ for $\theta=\{40,50,60\}$
are within QMC errorbars of each other. This shows that a projection
length of $\theta=40$ is sufficient for ground state values.
For $m^2$ (third panel of Fig. \ref{fig:proj}),
a similar conclusion holds.
A minor thing of note that can be seen in first panel of Fig. \ref{fig:proj} is
 that in comparison to
 the full range of $R \in [0,1]$, the full $\theta$ dependence
 for a given system size $\lesssim 5 \%$ which suggests that the 
 scenario
described in the main text will be applicable to finite temperatures
as well.

\onecolumngrid

%\begin{thebibliography}{999}

%\bibitem{Zanchi_Schulz}
%D. Zanchi and H.J. Schulz, Europhys. Lett. \textbf{44}, 235 (1997); D.
%Zanchi and H.J. Schulz, Phys. Rev. B \textbf{61}, 13609 (2000).

%\bibitem{Halboth_Metzner}
%C.J. Halboth and W. Metzner, Phys. Rev. B \textbf{61}, 7364 (2000);
%Phys. Rev. Lett. \textbf{85}, 5162 (2000).

%\bibitem{Honerkamp_Salmhofer}
%C. Honerkamp and M. Salmhofer, Phys. Rev. B \textbf{64}
%184516 (2001); Phys. Rev. Lett. \textbf{87}, 187004 (2001).

%\bibitem{strong}
%In the strong coupling phase,
% projection to the ground states is even faster
%because of finite single particle gap due to
%presence of (N\'eel)
%order.

\end{bibunit}

\end{document}